\documentclass[preprint,aps ,nofootinbib]{revtex4}
\usepackage{graphicx}
\usepackage{amsmath}
\usepackage{amsfonts}
\usepackage{amssymb}
\usepackage{color}%
\usepackage{dcolumn}
\newcommand{\f}{\begin{equation}}
\newcommand{\ff}{\end{equation}}
\newcommand{\fa}{\begin{eqnarray}}
\newcommand{\ffa}{\end{eqnarray}}

\begin{document}
\title{Holographic superconductors from the massive gravity}
\author{Hua Bi Zeng} \email{zenghbi@gmail.com}
\author{Jian-Pin Wu} \email{jianpinwu@gmail.com}
\affiliation{Department of Physics, School of Mathematics and Physics, Bohai University, Jinzhou, 121013, China}

\begin{abstract}

A holographic superconductor is constructed in the background of a
massive gravity theory. In the normal state without condensation,
the conductivity exhibits a Drude peak that approaches a delta function in the massless gravity limit
as studied by David Vegh.
In the superconducting state, besides the infinite DC conductivity, the AC conductivity has Drude behavior at low frequency followed by a power law-fall.
These results are in agreement with that found earlier by Horowitz and Santos, who studied a holographic
superconductor with an implicit periodic potential beyond the probe limit. The results also agree with measurements
on some cuprates.

\end{abstract} \maketitle

\section{Introduction}

As a new method to study strongly coupled field theory, the AdS/CFT correspondence\cite{adscft} has been used to study the quantum
many-body system in the past few years, which we call the AdS/CMT. A remarkable progress of AdS/CMT is the first holographic realization superconductivity\cite{h1},
for review see\cite{h2}. However, in this setup the superconductor is homogeneous without broken translational symmetry,
as a result of the momentum being not dissipate, the DC conductivity is infinite even in the normal metal phase of the superconductor.
The translational symmetry breaking in AdS/CFT
has been studied by the perturbative method and Drude(-like) peaks were discovered\cite{4,5,6,7,8}.
Some other related studies, including the construction of the inhomogeneous black hole solution and the corresponding holographic superconductor,
by the perturbative method have also been obtained\cite{1107.36771207.2943}.
Furthermore, instead of the perturbative method, the authors in Refs.\cite{9,10,12} constructed some periodic gravitational background to simulate the lattice
by introducing a periodic scalar field or chemical potential on the boundary.
With Einstein-DeTurck method, they solve the coupled partial equations numerically and the universal non-Drude frequency
scalings with finite DC conductivity are disclosed.
The first holographic setup of a lattice superconductor
was studied by introducing a periodic charge density beyond probe limit in \cite{11}, and the expected infinite DC conductivity
is also found in the superconducting state.
Besides the way to solve partial differential equations with inhomogeneous sources as in\cite{9,10,11,12,12v1},
the use of a helical lattice, Q-lattice, or using spatially dependent but massless scalars can also enables us to break translation invariance while preserving homogeneity\cite{13,1311.3292}.
We can also break translational symmetry spontaneously by introducing some topological terms or some addition field in the action\cite{13v1}.
At the same time, the first example of a modulated phase in string theory was found in\cite{Jokelav1}.
Soon thereafter, they worked it out in D2-D8 model and also generalized the mechanism with magnetic fields and stuff both in D3-D7 and D2-D8 models
\cite{Jokelav2}.

Another mechanics to break translation symmetry is to introduce a mass term for the graviton as\cite{14}
\begin{eqnarray}\label{LImassiveGraviton}
\mathcal{L}_I=\sqrt{-g}m^2\delta g_{tx}\delta g^{tx},
\end{eqnarray}
which can be implemented in the massive gravity framework\cite{Massivev1,Massivev2}.
Due to diffeomorphism breaking, the conservation of the stress-energy tensor in the massive gravity is broken.
However, the massive gravity suffer from the instability problem due to the introduction of the mass term for the graviton.
To eliminate Boulware-Deser ghost, a non-linear massive gravity is constructed by introducing higher order interaction terms in the action\cite{1011.1232}.
In this framework, a charged black brane solution has been constructed and the conductivity has also been calculated\cite{14}.
A recent review of massive gravity can be found in \cite{Massivev3}.

In this paper, we are trying to construct a holographic superconductor in the massive gravity theory.
In the normal metal phase state, by choosing proper values of the gravity mass we reproduced exactly the results in \cite{14}, both the
Drude peak and power-law scaling behavior with the exponents $\gamma$ equal to $-2/3$ of the AC conductivity.
This power-law scaling are in agreement with measurements of the normal phase
of bismuth-based cuprates\cite{cuprates1} and also the results in\cite{9,10}.

In the superconducting state, besides the infinite DC conductivity, the authors also found a Drude behavior at low frequency followed by a power law fall-off. These results are in striking agreement with measurements
on cuprates \cite{cuprates2} and also the implicit lattice results with condensation in\cite{11}. However, in \cite{11} the author
found that the coefficient  of the power law is temperature independent, this is different from the holographic superconductor from massive gravity,
in which the the coefficient  of the power law is temperature dependent.

The organization of the paper is as follows. In Section \ref{Geometry} we gives the
model of the holographic superconductor and the superconducting phase transitions
are also studied.
In Section \ref{ConductivityN} the equations of motion for fluctuations are derived and the conductivity
of the normal phase is reviewed. The conductivity of the superconducting state is given in Section \ref{ConductivityS}.
Finally, the discussions and conclusions are given in Section \ref{Conclusion}.

\section{Superconducting from non-linear massive gravity}\label{Geometry}

\subsection{Model and equations}
We start with the non-linear massive gravity action including a cosmological constant following \cite{14}
\begin{eqnarray}
\label{MassiveGravityAction}
S_{G}=\frac{1}{2\kappa^{2}}\int d^{4}x \sqrt{-g}
\left[R+\Lambda
+m_{g}^{2}\Sigma_{i=1}^{4}c_{i}\mathcal{U}_{i}(g,f)\right].
\end{eqnarray}
where $\Lambda=\frac{6}{L^{2}}$, $f$ is the reference metric, $c_{i}$ are constants and $m_g$ is the mass of the graviton.
$\mathcal{U}_{i}$ are symmetric polynomials of the eigenvalues of the matrix $\mathcal{K}^{\mu}_{\ \nu}\equiv \sqrt{g^{\mu\alpha}f_{\alpha\nu}}$
\begin{eqnarray}
\label{MathcalU}
\nonumber
&&
\mathcal{U}_{1}=[\mathcal{K}],
\\
\nonumber
&&
\mathcal{U}_{2}=[\mathcal{K}]^{2}-[\mathcal{K}^{2}],
\\
\nonumber
&&
\mathcal{U}_{3}=[\mathcal{K}]^{3}-3[\mathcal{K}][\mathcal{K}^{2}]+2[\mathcal{K}^{3}],
\\
&&
\mathcal{U}_{4}=[\mathcal{K}]^{4}-6[\mathcal{K}^{2}][\mathcal{K}]^{2}+8[\mathcal{K}^{3}][\mathcal{K}]
+3[\mathcal{K}^{2}]^{2}-6[\mathcal{K}^{4}],
\end{eqnarray}
where $[\mathcal{K}]\equiv \mathcal{K}^{\mu}_{\ \mu}$, $[\mathcal{K}^{2}]\equiv (\mathcal{K}^{2})^{\mu}_{\ \mu}$
and $(\mathcal{K}^{2})_{\mu\nu}\equiv \mathcal{K}_{\mu\alpha}\mathcal{K}^{\alpha}_{\ \nu}$ et. al..
As Ref.\cite{14}, we are interested in the case of a
spatial reference metric (in the basis $(t,z,x,y)$)
\begin{eqnarray}
\label{SpatialReferenceMetric}
f_{\mu\nu}=(f_{sp})_{\mu\nu}=diag(0,0,F^{2},F^{2}),
\end{eqnarray}
where $F$ is constant. In this case, there are only two spatial graviton mass terms
\footnote{The terms $\mathcal{U}_{3}$ and $\mathcal{U}_{4}$ vanish when we take the spatial reference metric (\ref{SpatialReferenceMetric}).}
\begin{eqnarray}
m_g^{2}\mathcal{U}_{sp}=m_g^{2}(\alpha\mathcal{V}_{1}+\beta\mathcal{V}_{2}),
\end{eqnarray}
where
\begin{eqnarray}
&&
\mathcal{V}_{1}=\mathcal{U}_{1}=[\mathcal{K}]=F(\sqrt{g^{xx}}+\sqrt{g^{yy}}),
\nonumber
\\
&&
\mathcal{V}_{2}=\mathcal{U}_{2}=[\mathcal{K}]^{2}-[\mathcal{K}^{2}]=2F^2\sqrt{g^{xx}g^{yy}},
\end{eqnarray}
and we have denoted $\alpha=c_1$, $\beta=c_2$. Without loss of generality, we will set $F=1$ in the following.

In order to built a holographic superconductor from non-linear massive gravity, we introduce a $U(1)$ gauge field and a charged complex scalar field, with action
\begin{eqnarray}
\label{MatterAction}
S_{M}=-\frac{1}{2\kappa^{2}}\int d^{4}x \sqrt{-g}\left(\frac{L^{2}}{4}\mathcal{F}_{\mu\nu}\mathcal{F}^{\mu\nu}+|D \Psi|^{2} + \frac{m_{\Psi}^{2}}{L^{2}} |\Psi|^{2}\right),
\end{eqnarray}
where $\mathcal{F}_{\mu\nu}=\partial_{\mu}A_{\nu}-\partial_{\nu}A_{\mu}$
and $D_{\mu}=\nabla_{\mu}-i q A_{\mu}$ with $m_{\Psi}$ and $q$ being the mass and charge of the complex scalar field, respectively.
For convenience, we will set $L=1$.
From the non-linear massive gravity action (\ref{MassiveGravityAction}) plus the matter action (\ref{MatterAction}),
one can obtain Einstein's equation, scalar equation and Maxwell's equation as, respectively
\begin{eqnarray}
\label{EinsteinEquation}
\nonumber
&&
R_{\mu\nu}-\frac{1}{2}g_{\mu\nu}R-\frac{1}{2}g_{\mu\nu}\Lambda+m_{g}^{2}X_{\mu\nu}
-\frac{1}{2}g^{\rho\sigma}\mathcal{F}_{\mu\rho}\mathcal{F}_{\nu\sigma}+\frac{1}{8}g_{\mu\nu}\mathcal{F}_{\rho\sigma}\mathcal{F}^{\rho\sigma}
\\
\
&&
-\frac{1}{2}(D_{\mu}\Psi D^{\ast}_{\nu}\Psi^{\ast}+D_{\nu}\Psi D^{\ast}_{\mu}\Psi^{\ast})
+\frac{g_{\mu\nu}}{2}(m_{\Psi}^{2}|\Psi|^{2}+|D\Psi|^{2})
=0,
\label{ScalarEquation}
\\
\
&&
D_{\mu}D^{\mu}\Psi-m_{\Psi}^{2}\Psi=0,
\label{MaxwellEquation}
\\
\
&&
\nabla^{\mu}\mathcal{F}_{\mu\nu}-iq[\Psi^{\ast}D_{\nu}\Psi-\Psi D_{\nu}^{\ast}\Psi^{\ast}]=0,
\end{eqnarray}
where
\begin{eqnarray}
\label{Xmunu}
X_{\mu\nu}=\frac{\alpha}{2}\left(\mathcal{K}_{\mu\nu}-[\mathcal{K}]g_{\mu\nu}\right)
-\beta\left( (\mathcal{K}^{2})_{\mu\nu}-[\mathcal{K}]\mathcal{K}_{\mu\nu} +\frac{1}{2}g_{\mu\nu}([\mathcal{K}]^{2}-[\mathcal{K}^{2}])\right).
\end{eqnarray}

We take the metric ansatz
\begin{eqnarray}
\label{Metric}
ds^{2}=\frac{L^{2}}{z^{2}}\left(-g(z)e^{-\chi(z)}dt^{2}+\frac{dz^{2}}{g(z)}+dx^{2}+dy^{2}\right),
\end{eqnarray}
with
\begin{eqnarray}
\label{At}
A=\Phi(z)dt,~~~~~~\Psi=\Psi(z).
\end{eqnarray}
From the $z$ component of Maxwell's equations, one finds that the
phase of the complex scalar field $\Psi$ must be constant so that we can set $\Psi$ being real without loss of generality.
The equations of the scalar field and the Maxwell field are, respectively
\begin{eqnarray}
\label{ScalarEquationCompenent}
&&
\Psi''+\left( \frac{g'}{g} - \frac{\chi'}{2} - \frac{2}{z}\right)\Psi'
+\frac{q^{2}e^{\chi}\Phi^{2}\Psi}{g^{2}}-\frac{m_{\Psi}^{2}\Psi}{z^2g}=0,
\label{MaxwellEquationCompenent}
\\
\
&&
\Phi''+\frac{\chi'\Phi'}{2}-\frac{2 q^{2} \Psi^{2}\Phi}{z^2g}=0.
\end{eqnarray}
Einstein's equations are
\begin{eqnarray}
\label{EinsteinEquation1}
&&
\Psi'^{2}-\frac{\chi'}{z}+\frac{q^{2}e^{\chi}\Phi^{2}\Psi^{2}}{g^{2}}=0,
\label{EinsteinEquation2}
\\
\
&&
\frac{3}{z^2}-\frac{3}{z^2g}-\frac{\alpha m^2_g}{zg}-\frac{\beta m^2_g }{g}
+\frac{m^2_\Psi \Psi^2}{2z^2g}+\frac{q^2e^{\chi}\Psi^2\Phi^2}{2g^2}-\frac{g'}{zg}+\frac{e^\chi z^2 \Phi'^2}{4g}+\frac{1}{2}\Psi'^2=0,
\end{eqnarray}

At the horizon $z_{+}$ ($g(z_+)=0$), by using a serious solution we can find that there are are independent parameters
\begin{equation}
z_{+}, \Psi_{+} \equiv \Psi(z_{+}), E_{+}\equiv \Phi'(z_{+}),\chi_{+}= \chi(z_{+}).
\end{equation}
The scalar
potential itself must go to zero at the horizon in order for the gauge connection to
be regular.
The Hawking temperature of the black solution is determined by the above quantities
\begin{equation}
T=\frac{1}{16 \pi L^2}((12+4 \Psi_{+}^2)e^{-\chi_{+}/2}-L^2 E_{+}^2 e^{\chi_{+}/2}+4(\alpha+\beta)e^{\chi_{+}/2}).
\end{equation}

By choosing a value of the scalar fields $m_\Psi^2=-2$, the information of the boundary field theory can be read from the asympototic behavior of
the fields at boundary $z=0$,
\begin{equation}
\Phi(z)=\mu-\rho z+\cdots,
\end{equation}
\begin{equation}
\Psi(z)=\Psi_1 z -\Psi_2 z^2+\cdots,
\end{equation}
where the $\mu$ is the chemical potential and $\rho$ is charge density on the boundary field theory.
The order parameter of the holographic superconductor is expectation value of the scalar operator,
it can be $\Psi_1$ or $\Psi_2$ depending on the quantization we choose.
One is to set $\Psi_1=0$ while $\Psi_2=\langle \mathcal O_2 \rangle$ is the order parameter.
The other is to treat $\Psi_2=0$ as the source while $\Psi_1=\langle \mathcal O_1 \rangle$ is the condensation value.

Before solving the four equations we can set both $L$ and $z_+$ equal to one  by using a scaling symmetry of the system.
With fixed $z_+=1$ we still have three independent parameter at the horizon. We should also fix the value of $\chi_+=0$
at the horizon, then after a rescaling of time $t\rightarrow ct$ with $c=e^{\chi(z=0)}$ we can set $\chi=0$ on the boundary.
Then the Hawking temperature of the black hole is to be the temperature of the
boundary field theory.

Finally, with the two independent parameter $\Psi_+$ and $\Phi_+$, we can solve the equations of motion by the shooting method after choosing a specific quantization for scalar field $\Psi$. We will see that after reducing  temperature the system will enter a superconducting state.

\subsection{the superconducting phase transition}

\begin{figure}
\begin{center}
\includegraphics[trim=0cm 0cm 0cm 0cm, clip=true,scale=0.7]{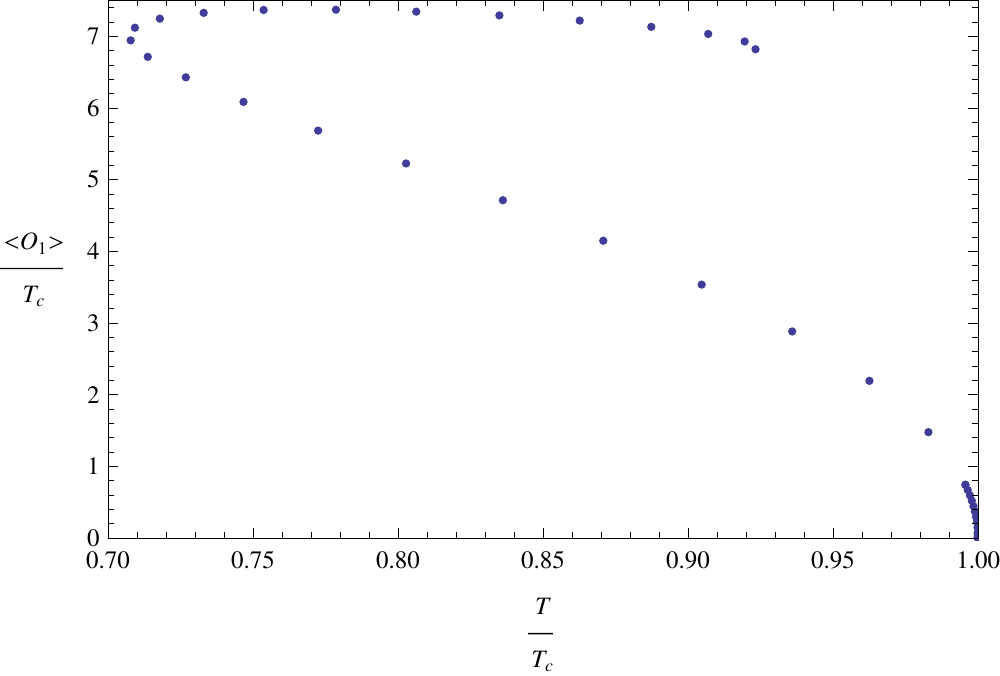}
\includegraphics[trim=0cm 0cm 0cm 0cm, clip=true,scale=0.7]{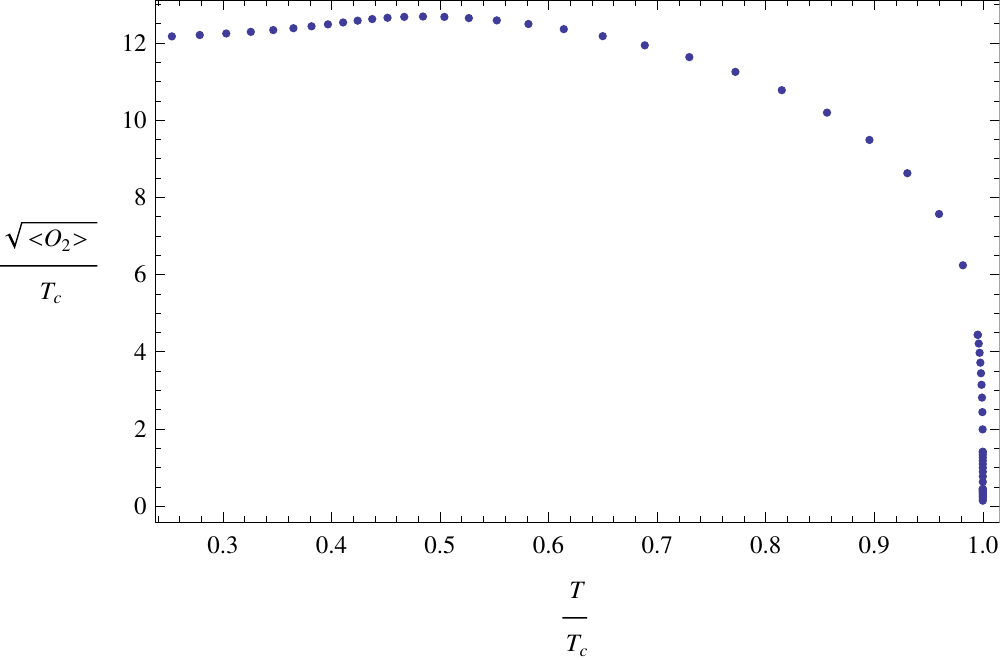}
\caption{\label{O1}The condensation phase diagram when $\alpha=-1,\beta=0, q=1$. }\label{fig1}
\end{center}
\end{figure}

\begin{figure}
\begin{center}
\includegraphics[trim=0cm 0cm 0cm 0cm, clip=true,scale=0.7]{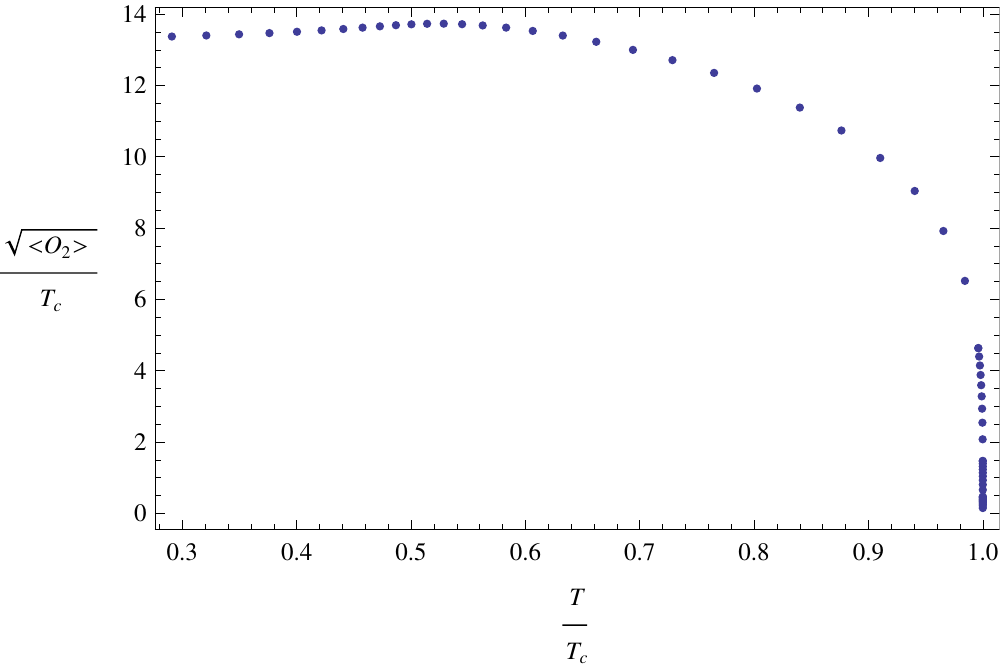}
\caption{\label{O2}The phase diagram when $\alpha=-0.75,\beta=0, q=1$. }\label{fig2}
\end{center}
\end{figure}

We choose two combinations  of $\alpha$ and $\beta$ as examples followed \cite{14}, which are $\alpha=-1,\beta=0$ and $\alpha=-0.75,\beta=0$.
The order parameter of  $q=1$ are shown in FIG.\ref{O1} and FIG.\ref{O2}. Note that in this paper, we will set $m_g=1$ throughout.

From the two figures we see that there indeed a critical temperature
below that there are condensation for $\langle \mathcal O_2 \rangle$.
However, the case of $\langle \mathcal O_1 \rangle$ is different,
for the parameter $\alpha=-0.75,\beta=0$ we even can not find a finite temperature solution
with condensation, when $\alpha=-1,\beta=0$, the condensation  $\langle \mathcal O_1 \rangle$
takes a two valued  behavior when $T/T_c <0.95$, which means that the
phase transition structure is complex. However, the condensed region near the critical temperature
is single valued, we choose this region when we compute the conductivity for the $\langle \mathcal O_1 \rangle$ case.
Numerical fitting of the data close to $T_c$ also indicates that the order parameter ~ $(1-T/T_c)^{1/2}$ at the critical point,
which admit the mean field exponent.



\section{Conductivity of normal state}\label{ConductivityN}

When there is no condensation, it has been confirmed in \cite{14,15} that for
a region of $\alpha$ and $\beta$, the conductivity takes a Drude like peak at
low frequency $\omega<T$, while $T<\omega<\mu$, the conductivity has a Power law
scaling $|\sigma| \approx C +\frac{B}{\omega^{2/3}}$ , $C$ and $B$ are
determined by $\alpha$ and $\beta$. In \cite{15}, the author analytically
computed the AC conductivity when the frequency is small and get exactly the
Drude model form.  In \cite{16}, a universal formula for the DC conductivity was
derived based on the argument of \cite{17} that the DC conductivity does not
vary with the radial coordinate, but depends only on geometric properties of the
horizon. This formula derived in \cite{16} relates the resistivity of the boundary field theory to the mass of the graviton evaluated on
the horizon of the bulk black hole as
\begin{equation}
\sigma_{DC}=\frac{1}{q^2}\left[1+\frac{q^2 \mu^2}{m^2(z_h)}\right],
\end{equation}
in which $m^2(z_h)=-2 \beta- \alpha/z_h$.
By choosing the same value of $\alpha$ and $\beta$ used in the
previous section we reproduce both the Drude like scaling
and the power-law scaling. The $\sigma_{DC}$ can also
vanish if we choose $q=1, \mu=1, \alpha=0,\beta=0.5$,
this maybe a holographic realization of coherent metal in massive gravity.

\subsection{Fluctuations and equations of motion}
The conductivity is computed from the retarded Green’s function of the transverse
current via the Kubo formula
\begin{equation}
\sigma(\omega)=\frac{i}{\omega} G^R(\omega).
\end{equation}

According to the $AdS/CFT$ correspondence,the Green function $G^R(\omega)$ is computed by solving the
equations of motion for fluctuations $a_x$ and the fields
coupled to it. No matter the temperature is below $T_c$ or above $T_c$, there are
three independent fluctuation fields $a_x, h_{tx}$ and $h_{zx}$. The linear equations
of the three fields are

\begin{eqnarray}
\label{EMx}
&&
a''_{x}+\left(\frac{g'}{g}-\frac{\chi'}{2}\right)a'_{x}
+\left(\frac{\omega^{2}e^{\chi}}{g^{2}}-\frac{2q^{2}\Psi^{2}}{z^{2}g}\right)a_{x}
+\frac{z^{2}e^{\chi}\Phi'}{g}\left(\frac{2}{z}h_{tx}+h'_{tx}-i\omega h_{zx}\right)=0,
\label{EStx}
\\
&&
(\frac{2}{z}h_{tx}+h_{tx}'-i \omega h_{zx}+\Phi' a_x)=-\frac{i e^{-\chi} g m^2(z)}{\omega} h_{zx},
\label{ESrx}
\\
&&
(\frac{2}{z}h_{tx}+h_{tx}'-i \omega h_{zx}+\Phi' a_x)'=\frac{m^2(z)}{g}h_{tx}-\frac{1}{2}\chi'(\frac{2}{z}h_{tx}+h_{tx}'-i \omega h_{zx}+\Phi' a_x).
\end{eqnarray}
in which $m^2(z)=-2 \beta -\frac{\alpha}{z}$.
If there is no condensation, the equations above will come back to the equations
studied in \cite{14,15}.

We can express $h_{tx}$ as a function of $h_{zx}$,
\begin{equation}
\frac{g}{m^2(z)}\left[-\frac{i e^{-\chi}g m^2(z) }{\omega} h_{zx}\right]'+(-\frac{i e^{-\chi} \chi' g^2 }{2\omega} h_{zx})=h_{tx}.
\end{equation}

Finally we get two coupled non-linear equations for $a_x$ and $h_{zx}$,
\begin{equation}
a''_{x}+\left(\frac{g'}{g}-\frac{\chi'}{2}\right)a'_{x}
+\left(\frac{\omega^{2}e^{\chi}}{g^{2}}-\frac{2q^{2}\Psi^{2}}{z^{2}g}\right)a_{x}
+\frac{z^{2}e^{\chi}\Phi'}{g}\left(-\Phi' a_x-\frac{i e^{-\chi} g m^2(z)}{\omega}h_{zx}\right)=0,
\end{equation}
\begin{eqnarray}
&&
\nonumber
\frac{2}{z}\left[\frac{g}{m^2(z)}\left(-\frac{i e^{-\chi}g m^2(z) }{\omega} h_{zx}\right)'+\left(-\frac{i e^{-\chi} \chi' g^2 }{2\omega} h_{zx}\right)\right]
\\
\nonumber
&&
+\left[\frac{g}{m^2(z)}\left(-\frac{i e^{-\chi}g m^2(z) }{\omega} h_{zx}\right)'\right]'+\left(-\frac{i e^{-\chi} \chi' g^2 }{2\omega} h_{zx}\right)'
\\
&&
-i \omega h_{zx}+\Phi' a_x=-\frac{i e^{-\chi} g m^2(z)}{\omega} h_{zx}.
\end{eqnarray}
The equations of motion is solved numerically by setting a ingoing boundary condition
at the horizon that
\begin{equation}
a_x(z),h_{zx}(z)\propto (1-z)^{-\frac{ i \omega}{ 4 \pi T }}.
\end{equation}
We set normalizable UV boundary conditions for the $h_{zx}$
field. On the boundary, the behavior of the $a_x$ is
\begin{equation}
a_x(z)=a_1+ a_2 z +\cdots.
\end{equation}
And then the retard green function is read as
\begin{equation}
G^R(\omega)=\frac{a_2}{a_1},
\end{equation}
Therefore, the conductivity can be expressed as
\begin{equation}
\sigma(\omega)=\frac{i}{\omega}\frac{a_2}{a_1}.
\end{equation}

In the next subsection, we will present the results of conductivity with both Drude and power law scaling.

\subsection{Drude like scaling and pwer law scaling}

\begin{figure}
\begin{center}
\includegraphics[trim=0cm 0cm 0cm 0cm, clip=true,scale=0.7]{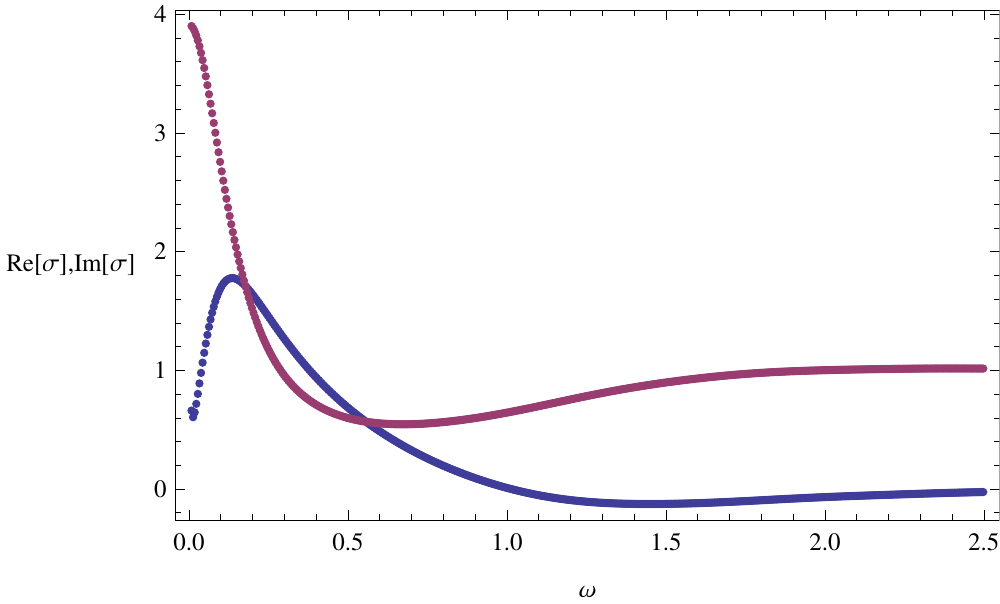}
\includegraphics[trim=0cm 0cm 0cm 0cm, clip=true,scale=0.7]{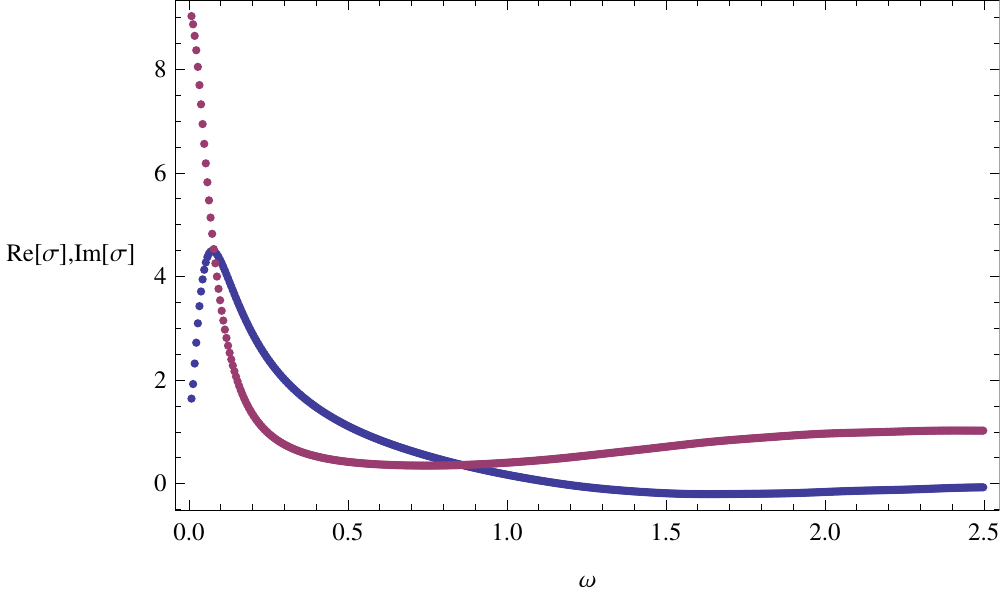}
\caption{Both real part and imaginary part of conductivity. Left: $\alpha=-1,\beta=0,\mu=1.732$. Right: $\alpha=-0.75,\beta=0,\mu=2.5$.
We can see there is a Drude peak at low frequency.}\label{fig3}
\end{center}
\end{figure}

\begin{figure}
\begin{center}
\includegraphics[trim=0cm 0cm 0cm 0cm, clip=true,scale=0.7]{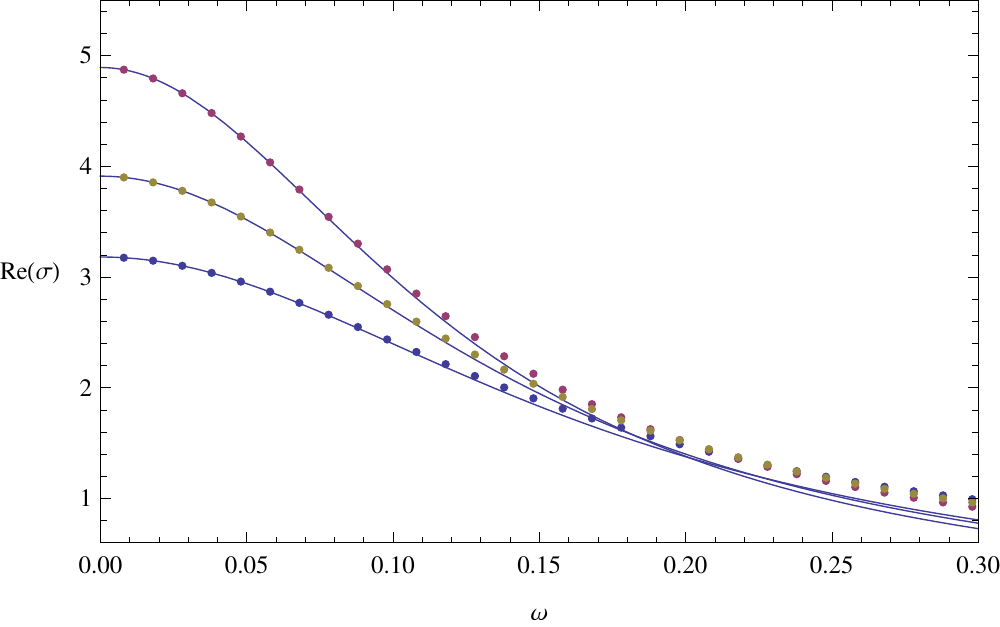}
\includegraphics[trim=0cm 0cm 0cm 0cm, clip=true,scale=0.7]{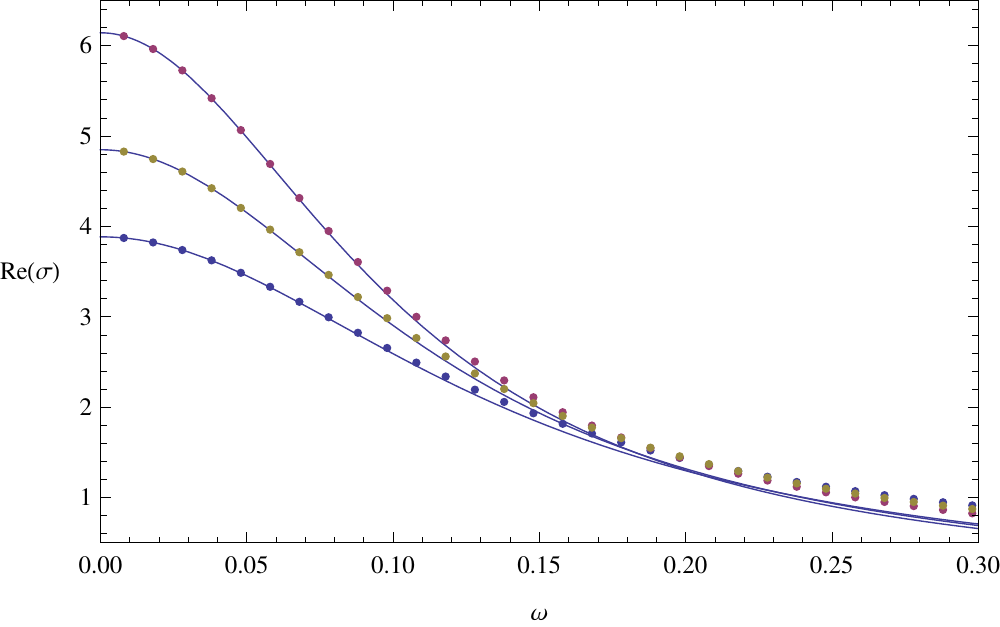}
\caption{ Left: The Drude scaling for three different
temperature $T=0.08;0.1;0.114$ (from top to bottom) for $\alpha=-1,\beta=0$. Right: The Drude scaling for three different
temperature $T=0.1;0.119;0.134$ (from top to bottom) for $\alpha=-0.75,\beta=0$.  When $\omega<T$, the
curves can be fitted very well with the Drude scaling $\sigma(\omega)= \sigma_{DC}/(1-i\omega \tau)$ when $\omega<T$. We also see that the $\sigma_{DC}$ decease
when increasing the temperature and finally goes to one. }\label{fig4}
\end{center}
\end{figure}

Following \cite{14}, we choose $\alpha=-1,\beta=0$ and $\alpha=-0.75,\beta=0$.
In FIG.\ref{fig3} the conductivity of the both cases above are presented, the Drude peak
is clear and the real part goes to one at large frequency.
The Drude scaling at low frequency are shown in FIG.\ref{fig4} for three
different temperature of both $\alpha=-1,\beta=0$ and $\alpha=-0.75,\beta=0$.
When $\omega<T$ the data can be fitted well very well by the
Drude formula
\begin{equation}
\sigma(\omega)=\frac{\sigma_{DC}}{1-i \omega \tau},
\end{equation}
in which $\tau$ is the relaxation time.

\begin{figure}
\begin{center}
\includegraphics[trim=0cm 0cm 0cm 0cm, clip=true,scale=0.7]{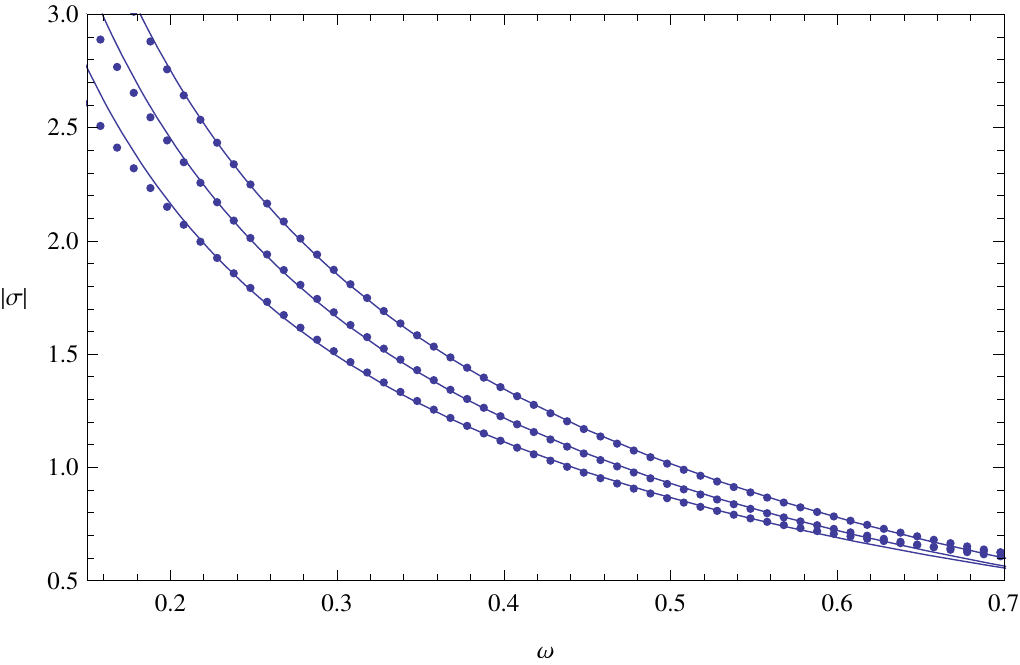}
\includegraphics[trim=0cm 0cm 0cm 0cm, clip=true,scale=0.75]{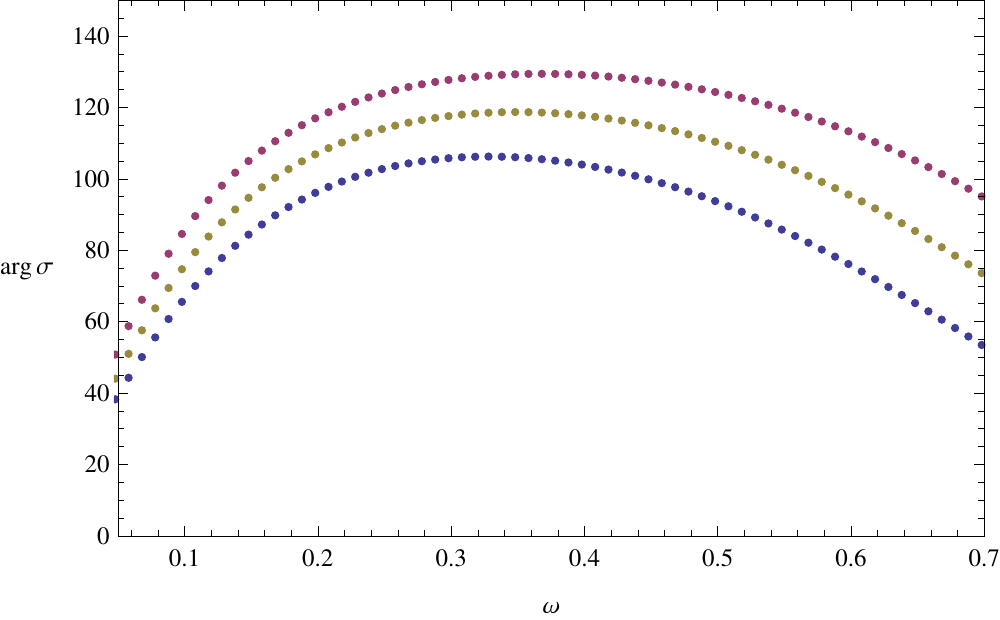}
\caption{\label{FIGPowerLaw}Left: The power law scaling for three different
temperature $T=0.1;0.119;0.134$ (from top to bottom) for $\alpha=-0.75,\beta=0$. The plots are
fitted very well by $|\sigma(\omega)|= C+ B \omega^{-2/3}$. Right: The phase of $\sigma(\omega)$. }
\end{center}
\end{figure}

In the intermediate frequency the absolute value of the  conductivity can be fitted well by the power law scaling
\begin{equation}\label{PowerLaw}
|\sigma(\omega)|=C+ B\omega^{-\gamma}.
\end{equation}
By a fine-tuning of the graviton mass, we can obtain $\gamma\approx 2/3$ (FIG.\ref{FIGPowerLaw}), as claimed in the lattice models\cite{9,10,11,12}.
In addition, the results of $\alpha=-0.75,\beta=0$ for three different
temperature have also been presented in FIG.\ref{FIGPowerLaw}, which indicate that the power law scaling hold
when changing temperature. However, we must also point out that for a certain class of holographic lattice models, the so-called holographic \emph{Q-lattice} model\cite{1311.3292},
the power law scaling of $2/3$ exponent found in \cite{9,10,11} is not universal,  which may be depend on the specific model and need to be furthermore studied.

Both the Drude scaling at low frequency $\omega<T$ and Power law scaling at
an intermediate regime $T < \omega <\mu$  shown above are in agreement with the results
in \cite{9,10,12} when an explicit inhomogeneous
lattice is introduced in the bulk theory. There must be connections
between the massive gravity theory and the inhomogeneous lattice models,
in \cite{18}, by using the small-lattice
expansion, the authors proved that, the presence of the latticed
scalar source induces an effective mass for the graviton via a gravitational version of the Higgs
mechanism.




\section{Conductivity of superconducting state}\label{ConductivityS}

In this section we are going to discuss the conductivity in the condensation
state with non-zero $\Psi$. As a results of superconducting we see that the
DC conductivity will become infinite. However, the finite frequency AC conductivity
admit the same scaling behavior as the normal state as discussed in the previous
section. Especially the power law scaling with $\gamma \approx -2/3$
are in agreement with the measurement in cuprates\cite{cuprates1,cuprates2}.

\begin{figure}
\begin{center}
\includegraphics[trim=0cm 0cm 0cm 0cm, clip=true,scale=0.7]{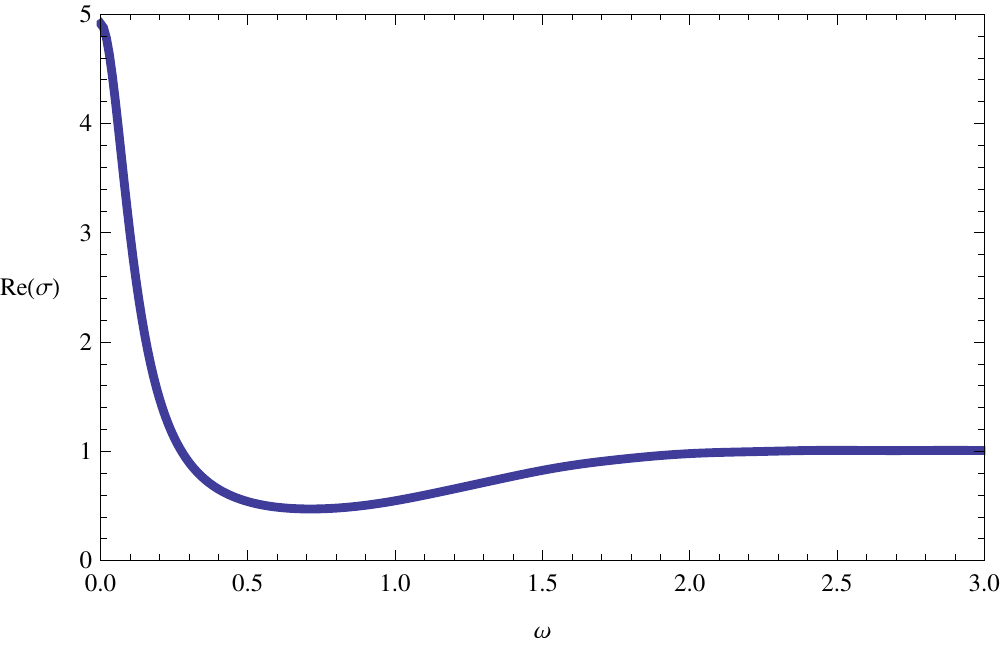}
\includegraphics[trim=0cm 0cm 0cm 0cm, clip=true,scale=0.7]{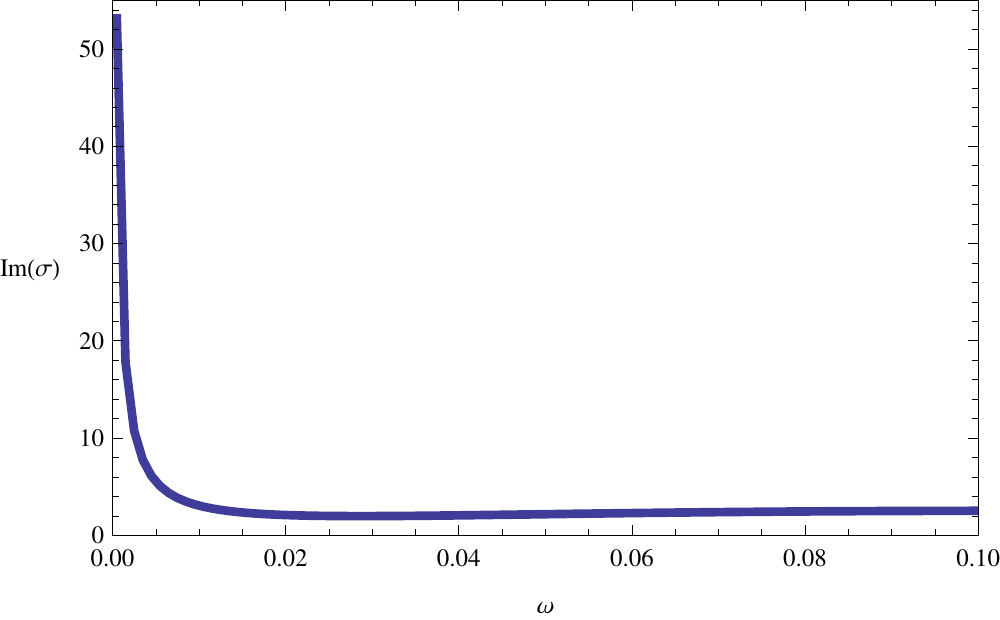}
\caption{The conductivity of the condensation phase with $\langle \mathcal O_1 \rangle$ for $\alpha=-1,\beta=0$ and $T/T_c=0.99$.}\label{fig6}
\end{center}
\end{figure}

At temperature is $T/T_c=0.99$,
the conductivity of the condensation phase with $\langle \mathcal O_1 \rangle$ of $\alpha=-1,\beta=0$
are plotted in FIG.\ref{fig6}. Compared the results of normal state in previous section, the
difference is that the imaginary part of conductivity diverges at zero frequency.
The divergence of imaginary part when $\omega=0$ indicates that there is a delta function of the real part
conductivity, which confirms the exist of superconducting state.

\subsection{Low frequency region}

When $\omega$ is small, we find that the conductivity can be fitted well
by the formula as found in \cite{11}
\begin{equation}\label{sigma}
\sigma(\omega)=i \frac{\rho_s}{\omega}+\frac{\rho_n \tau}{1-i \omega \tau}.
\end{equation}
where $\rho_s$ is the superfluid density and  $\rho_n$ is the normal
fluid density, $\tau$ is the relaxation time.

\begin{figure}
\begin{center}
\includegraphics[trim=0cm 0cm 0cm 0cm, clip=true,scale=0.7]{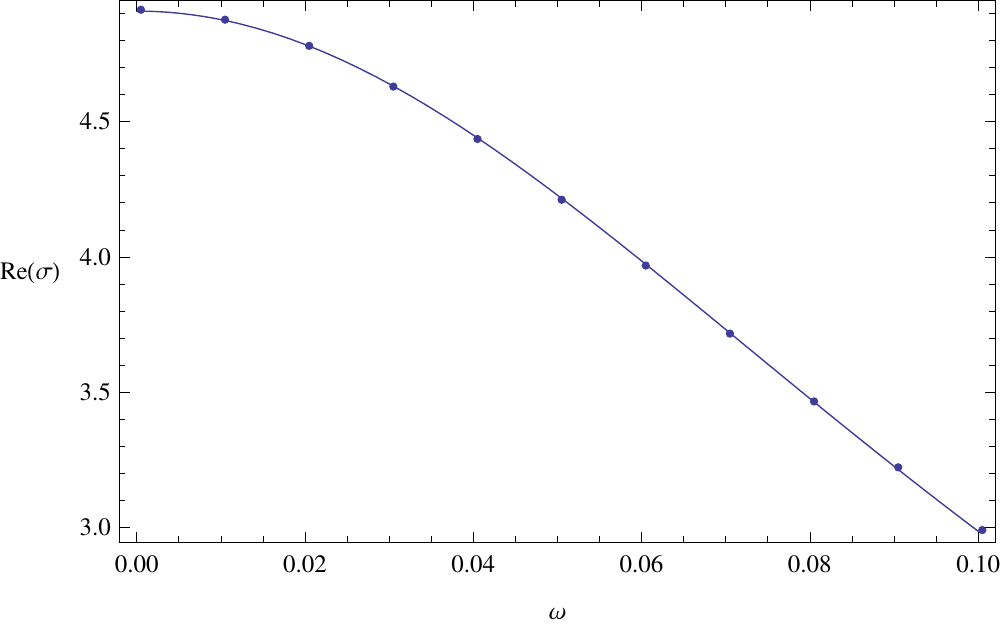}
\includegraphics[trim=0cm 0cm 0cm 0cm, clip=true,scale=0.7]{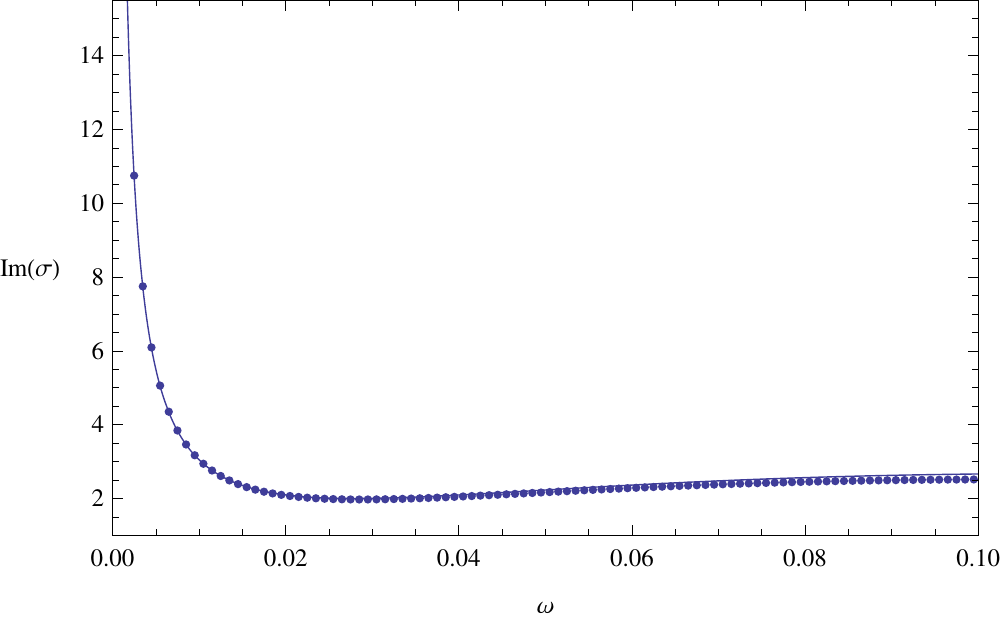}
\caption{At low frequency, the scaling of conductivity of the condensation phase with $\langle \mathcal O_1 \rangle$ for $\alpha=-1,\beta=0$ and $T/T_c=0.99$.
The plots can be fitted very well by Eq.(\ref{sigma}).}\label{fig7}
\end{center}
\end{figure}

\begin{figure}
\begin{center}
\includegraphics[trim=0cm 0cm 0cm 0cm, clip=true,scale=0.65]{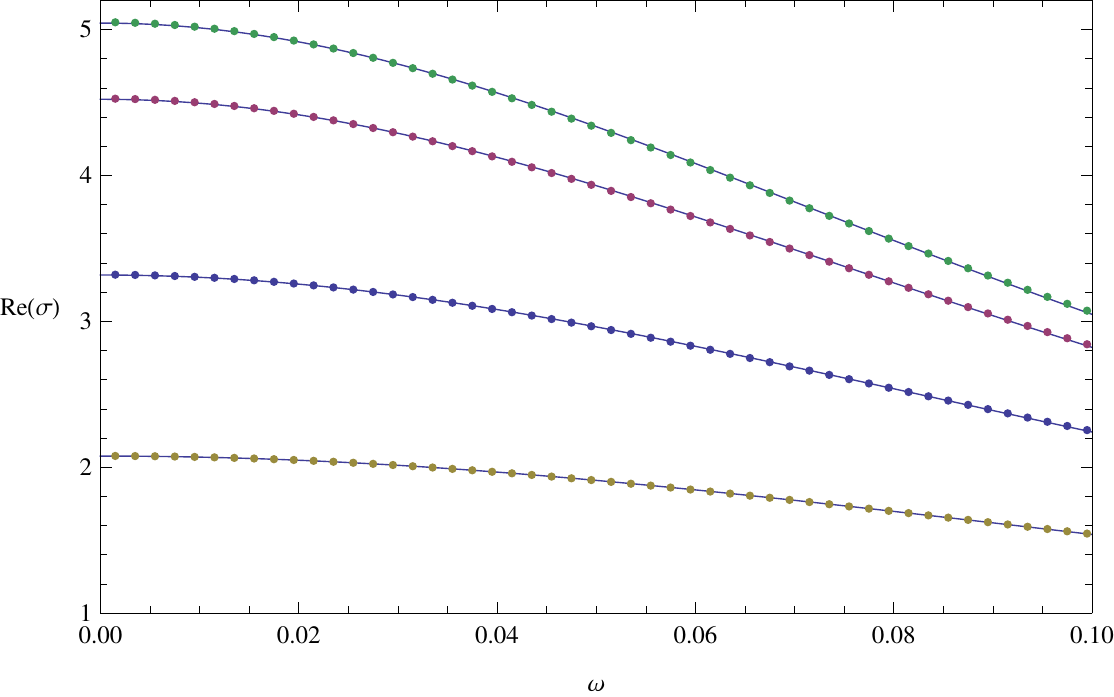}
\includegraphics[trim=0cm 0cm 0cm 0cm, clip=true,scale=0.7]{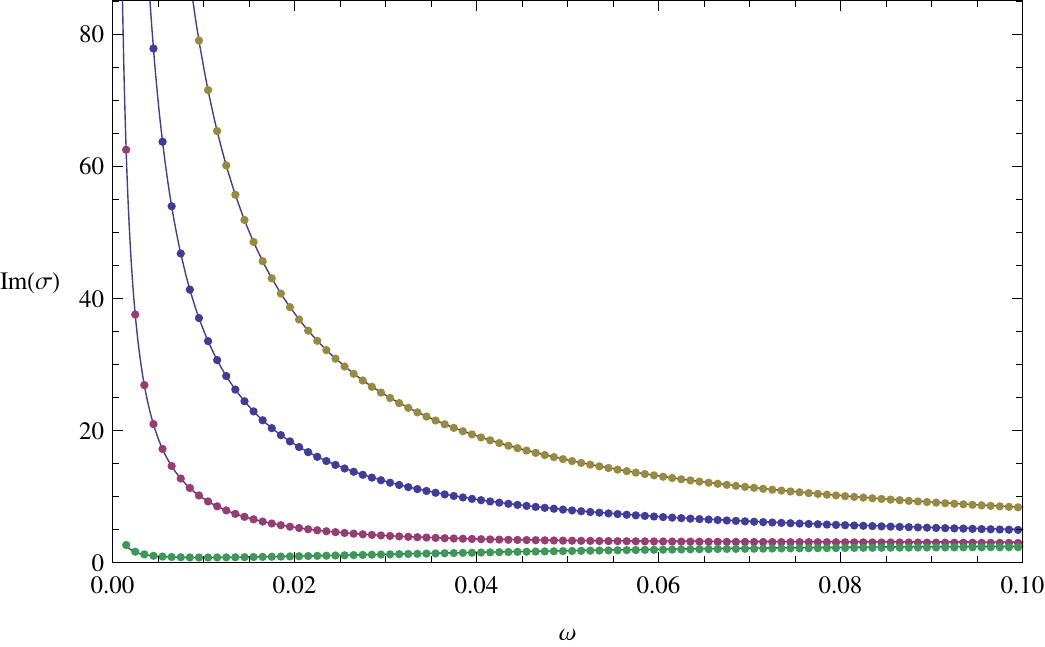}
\caption{$\alpha=-1,\beta=0$, at low frequency, scaling of conductivity of the condensation phase with $\langle \mathcal O_2 \rangle$ , the
temperature from top to to bottom is $T/T_c=0.999;0.981;0.930;0.856$. The data can be fitted very well by Eq.(\ref{sigma}).}\label{fig8}
\end{center}
\end{figure}

In FIG.\ref{fig7} and FIG.\ref{fig8}, we present the low frequency conductivity for both $\langle \mathcal O_1 \rangle$ and $\langle \mathcal O_2 \rangle$
of $\alpha=-1,\beta=0$. The agreement with Eq. (\ref{sigma}) is very convincing.
From FIG.\ref{fig8}, we also see that when the temperature is lowered, the pole of the imaginary conductivity grows rapidly
since there is more superfluid density $\rho_s$.

\begin{figure}
\begin{center}
\includegraphics[trim=0cm 0cm 0cm 0cm, clip=true,scale=0.6]{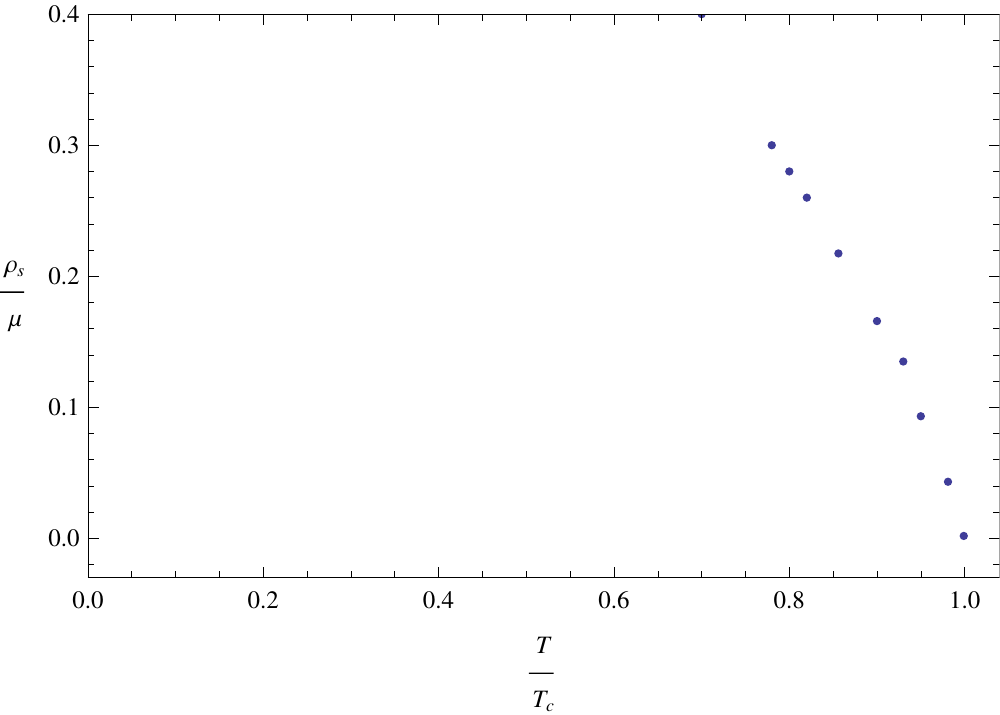}
\includegraphics[trim=0cm 0cm 0cm 0cm, clip=true,scale=0.7]{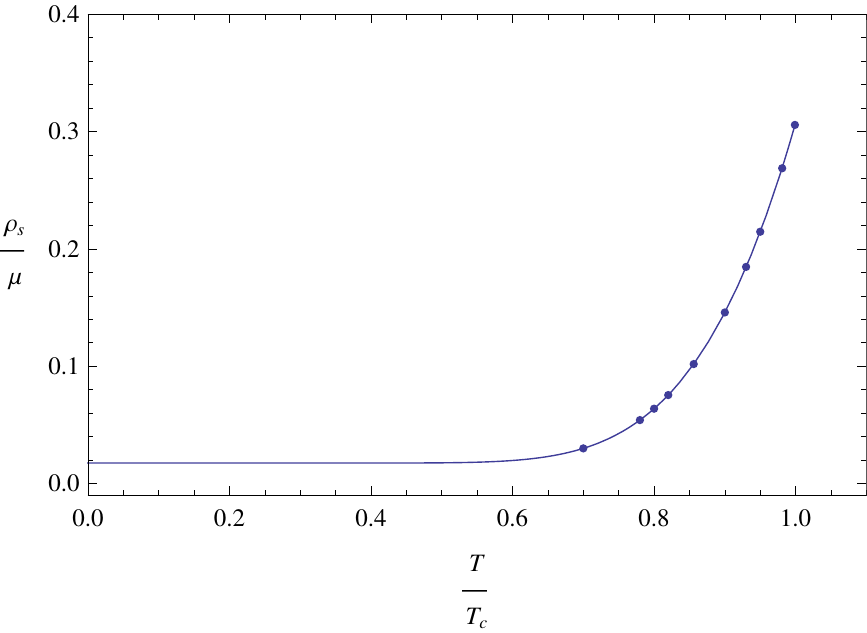}
\caption{The $\rho_n$ and $\rho_s$ in the superconducting state. }\label{fig9}
\end{center}
\end{figure}

The temperature dependence of both $\rho_n$ and $\rho_s$ are plotted in FIG.\ref{fig9},
we are working in the case  of  $\langle \mathcal O_2 \rangle$
when $\alpha=-1,\beta=0$. The $\rho_s$ is zero at $T_c$ and increases rapidly when lowing
temperature. The normal fluid density $\rho_n$ can be fitted very well as
\begin{equation}
\rho_{n}(T)= a + b e^{-\Delta/T}, \Delta \approx 6 T_c.
\end{equation}

This also similar to the BCS theory and also the result in \cite{11}.
The $\Delta \approx 6 T_c$ is much larger than the value of a BCS superconductor
with $\Delta \approx 1.7 T_c$ means that the holographic superconductor in
the massive gravity is indeed a strongly coupled superconductor.This
is also a significant distinction that highlights how holographic lattice
and massive gravity models generically lead to results that are
qualitatively similar but quantitatively different.

\subsection{Power law scaling and cuprates }

\begin{figure}
\begin{center}
\includegraphics[trim=0cm 0cm 0cm 0cm, clip=true,scale=0.8]{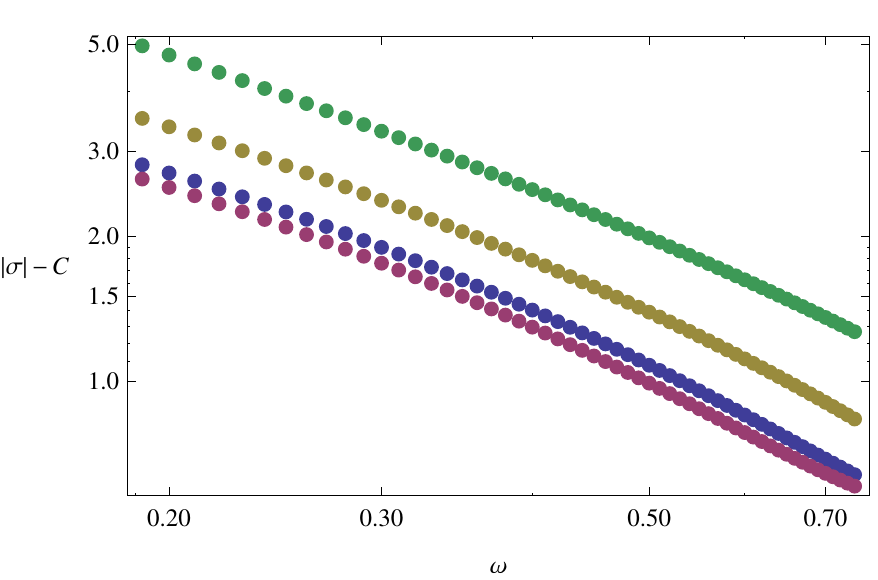}
\includegraphics[trim=0cm 0cm 0cm 0cm, clip=true,scale=0.8]{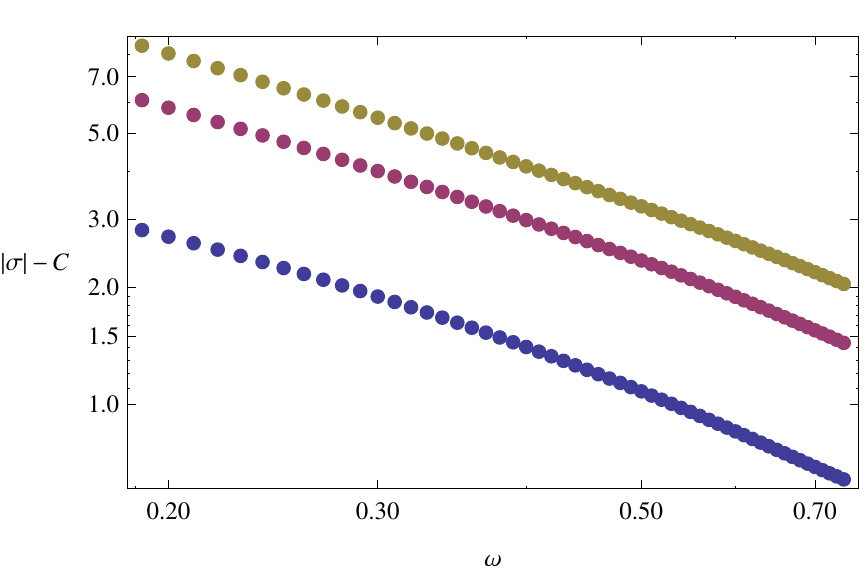}
\caption{The log-log plot of the normal part conductivity. Left: $\alpha=-1,\beta=0$, the condensation phase with $\langle \mathcal O_2 \rangle$ , the
temperature from top to to bottom is $T/T_c=0.999;0.981;0.930;0.856$. Right: from top to bottom are the cases: ($\alpha=-0.75,\beta=0,\langle \mathcal O_2 \rangle>0, T/T_c=0.98$); ($\alpha=-1,\beta=0,\langle \mathcal O_1 \rangle>0, T/T_c=0.98$) and ($\alpha=-1,\beta=0,\langle \mathcal O_2 \rangle>0,T/T_c=0.856$).  }\label{fig10}
\end{center}
\end{figure}

At slightly larger frequency, the absolute value of the conductivity of the normal component follows
the same power law Eq.(\ref{PowerLaw}) as was found in the normal phase above the critical temperature.
In FIG. 10  we  plotted the absolute value of the $\sigma(\omega)$ minus the off-set C vs
frequency on a log-log plot after subtracting the part from superfluid density. In the left side of FIG. \ref{fig10},
the four lines corresponding to the same temperature as studied in FIG. \ref{fig8} for the case $\alpha=-1,\beta=0$ with
condensation $\langle \mathcal O_2 \rangle$.
While in the right side we plotted the results of both  $\alpha=-1,\beta=0$ and $\alpha=-0.75,\beta=0$,
and also both two kinds of condensation $\langle \mathcal O_1 \rangle$ and $\langle \mathcal O_2 \rangle$. This clearly shows that the power law
fall-off is completely unaffected by the condensation.
Here, we also point out that we also need a fine-tuning of the graviton mass as that in the normal state.

The fact that the lines are parallel implies that the exponent is the same,
which is in agreement with the results in \cite{11} and also the experiment on
bismuth-based cuprates \cite{cuprates2}. The fact that the lines does not lie on
top of each other implies that the coefficient $B$ of the power law is temperature dependent,
this is different from the results in \cite{11}, in which the author found  the coefficient $B$ of the power law is temperature independent.

\section{conclusion}\label{Conclusion}
In this paper, a holographic superconductor with momentum
relaxation is constructed in the massive gravity proposed in
\cite{14}. In the normal state we reproduced the Drude scaling and
power-law scaling as studied in \cite{14,15}, while in the supercoduncting
state, the superconducting part induces a delta function for real part
conductivity when $\omega=0$. However, the Drude scaling and power-law
scaling of the normal part are not affected by the condensation.
The results agree with the result in \cite{11}, in which the holographic
superconductor with implicity lattice is studied.
This agreement confirm that there is a deep connection between the
massive gravity and the holographic lattice theory as studied in \cite{18}.
Recently, a holographic theory of strange metals is constructed from the Einstein-Maxwell-Dilaton
massive gravity\cite{19,20}, it would be interesting to find the superconductor/strange metal
phase transitions in this setup.
In this paper, we only focus on the cases of $\alpha$ and $\beta$ that produce the
normal metal state with Drude scaling, however, as pointed in \cite{16}, there are
other possibility that the conductivity will show possible incohenrent
metal behaviors by choosing a proper mass term, this is also worth a further studying.

\vspace{0.2in}   \centerline{\bf{Acknowledgments}} \vspace{0.2in} We are grateful to David Vegh and Yong Qiang Wang for helpful discussions.
HBZ is supported by the National Natural Science Foundation of China (under
Grant No. 11205020). J. P. Wu is supported
by the Natural Science Foundation of China under Grants
No. 11305018 and No. 11275208.

\end{document}